\documentclass[a4paper,10pt]{article}
\usepackage{amsfonts,amssymb,amsmath,bbm,bm}
\usepackage[utf8]{inputenc}
\usepackage[dvips]{graphicx}
\usepackage{hyperref}
\title{Spinor  representation of surfaces 
and complex stresses on membranes and interfaces}
\author{Jemal Guven\footnote{\href{mailto:jemal@nucleares.unam.mx}{jemal@nucleares.unam.mx}}\, and
Pablo V\'azquez-Montejo\footnote{\href{mailto:vazqmont@nucleares.unam.mx}{vazqmont@nucleares.unam.mx}}}
\date{}

\begin{document}

\maketitle

\begin{center}\it
Instituto de Ciencias Nucleares, Universidad Nacional Aut\'onoma de M\'exico,\\
Apdo. Postal 70-543, 04510 M\'exico D.F., Mexico
\end{center}

\begin{abstract}
Variational principles  
are developed within the framework of a spinor representation of the surface geometry
to examine  the equilibrium properties of a membrane or interface. 
This is a far-reaching generalization of the Weierstrass-Enneper 
representation for minimal surfaces, introduced by mathematicians in the nineties, 
permitting the relaxation of the vanishing mean curvature constraint. 
In this representation the surface geometry is described by  a spinor field,
satisfying a two-dimensional Dirac equation, 
coupled through  a potential associated with the mean curvature. As an application, 
the mesoscopic model for a fluid membrane as a surface described by the Canham-Helfrich energy 
quadratic in the mean curvature is examined. An explicit construction is provided of the  
conserved complex-valued stress tensor characterizing this surface. 
\end{abstract}

\section{Introduction}

The coarse-grained description of a physical surface is often entirely insensitive to microscopic
details, involving an energy function that depends only on the surface geometry. A familiar example
is provided by a soap film or an interface which will reach equilibrium when its
surface area is minimized. In the absence of a pressure difference across the surface, this state is
described by a minimal surface characterized by a vanishing mean curvature; it will thus assume a
saddle shape almost everywhere.

\vskip1pc\noindent
From a mathematical point of view minimal surfaces are very special: the functions embedding the
surface in space are harmonic so that, in an appropriate parametrization, the surfaces will be
described by analytic functions. This is the essential element of the classical Weierstrass-Enneper (WE)
representation of minimal surfaces, introduced 150 years ago, 
 which has played a central  role in the development 
of the subject ever since \cite{Weier,GrayFomenko}.  While physically it is often adequate to 
settle for a small gradient approximation in terms of a height function, where one can get by with a
more modest toolkit, the WE representation of the surface geometry provides an indispensable handle
on the global features of minimal surfaces; the Schwartz P Surface or plumber's nightmare
\cite{Schwarz}, describing liquid crystalline phases of fluid membranes, provides a good case in point
\cite{Tatea}.
 
\vskip1pc\noindent
Even if one is interested in interfaces, however, one almost always needs to accommodate a pressure
difference across this surface; the minimal surface is replaced by one of constant mean curvature.
Never mind addressing questions of stability or examining  fluctuations,
simple as this modification may be, its description lies beyond the scope of the WE representation.

\vskip1pc\noindent
There is, however, a generalization of the WE framework that applies to any surface. 
 Back in  1979 Kenmotsu showed how the classical representation could be tweaked
to describe surfaces with any prescribed mean curvature \cite{Kenmotsu}; a decade later, the WE
representation was reformulated in terms of a two-component spinor field, which also
turned out to be its natural setting \cite{Sul,KS1}. By the mid-nineties, this framework had been
extended  to accommodate a non-vanishing mean curvature \cite{Konop,KS2}. As 
emphasized by Konopelchenko and Taimanov in \cite{KNPTMV}, it was no longer even appropriate to
think of the mean curvature as prescribed. For reviews see references \cite {Bobenko} and \cite{Taimanov}.

\vskip1pc\noindent
The spinor appearing in this framework satisfies a linear Dirac-type equation on the complex plane;
its components couple through a real-valued external potential. As in the classical WE
representation, the functions which describe the embedding of the surface in space occur as
integrals of closed differential one-forms quadratic in the spinor field. 
The mean
curvature of the surface  is proportional to the potential. When it vanishes, the
components uncouple and the original representation for minimal surfaces is recovered.

\vskip1pc\noindent
Our focus will be on the mathematical description of fluid membranes on mesoscopic scales \cite{Seifert}.
On these scales the membrane is described with unusual accuracy as a two dimensional surface;
the energy associated with a given configuration is the Canham-Helfrich bending energy
quadratic in the mean curvature \cite{CanHel}. Unlike any familiar elastic material, in-plane shear
goes unpenalized; thus the surface is characterized completely by its geometrical degrees of 
freedom and membrane elasticity can be addressed in terms of geometry. A striking feature of
this energy  in the spinor representation 
is that the integrated mean curvature squared depends only of the potential.
 Modulo a topological contribution to the energy, this is also the most
general geometrical energy associated with bending. It is thus curious that
there has been no systematic attempt made to look more closely at this functional in terms of these
variables. 
In this paper we will examine the behavior of the Canham-Helfrich energy under  
variations of the potential; in particular we will show how the equation describing equilibrium 
shapes--non-linear both in the spinor and in the potential--is obtained. 

\vskip1pc\noindent
What would appear to be the obvious way to do this also turns out to be riven with technical
difficulties: if the spinors and the potential are treated as the fundamental variables, the
relationship connecting their variations to that of the embedding functions defining 
the surface is  non-local. One way to sidestep this difficulty is to introduce an appropriate set
of Lagrange multipliers, a spinor analogue of a framework introduced a few years ago to reformulate
the variational principles for parametrized surfaces in terms of the induced metric and the
extrinsic curvature \cite{auxil}. Just as the metric and the extrinsic curvature are not independent
variables, the spinor and the potential are also constrained (by the Dirac equation). This is just
as well; for a naive variation of the potential appears to imply that the only critical points of
bending energy are minimal surfaces, a manifestly incorrect conclusion. In this framework, the
relationship connecting the spinors to a surface, characterized by its three embedding functions,
will also be input  as a constraint. 
The multiplier enforcing this constraint  
quantifies the local change in energy as the surface is deformed; it is thus identified as a
``stress tensor''. Unlike the stress in a typical elastic medium, however, it will be determined
completely by the geometry. A surface in
mechanical equilibrium is  described by a conserved stress tensor. 

\vskip1pc\noindent
The description of equilibrium in terms of a geometrical stress tensor has been discussed previously in the 
context of a parametrized description of the surface; in reference \cite{Stress} it
was identified as the Noether current associated with the translational invariance of the energy,
an approach refined in reference \cite{auxil} using the framework of auxiliary variables. 
More physical continuum mechanical/thermodynamical treatments have been presented 
in references \cite{Steigmann} and \cite{Lomholt}).  Work in the latter direction dates back to 
the work of Evans \cite{Evans}.
The concept of geometrical surface stress is also familiar, if only implicitly, in the context of
minimal surfaces \cite{Hoffman,Kusner} where they appear in the identification of the
weights--external forces--that provide the tension preventing the collapse of the surface. 

\vskip1pc\noindent  
A problem of significant current biophysical interest is the study of interactions between 
surface-bound particles mediated by the surface geometry, with the particles in question identified as
proteins (see reference \cite{Oster}, for example, and more recently \cite{Steigmannproteins}). In
\cite{interactions} this problem was approached in the non-linear regime by examining the
geometrical stresses associated with the deformed geometry. By casting the Euler-Lagrange equations
as a conservation law it became possible to  gain access to global information
which is not manifest when the divergence is dismantled into tangential and normal parts.
Just as the techniques of  complex variables can be exploited very effectively in the linear regime (as demonstrated spectacularly in reference \cite{Oster}), the spinor framework  renders it possible to apply these same tools to examine physical processes which lie beyond the scope of a linear description. Recent reviews, which summarize nicely the role of geometrical stresses in a soft matter and biophysical
context, have been provided by Deserno \cite{Deserno} and Powers \cite{Powers}.

\vskip1pc\noindent
The simplification of the surface geometry  in  this representation involves giving
up, to a large degree, the reparametrization invariance inherent in the geometrical nature of the
problem. An
arbitrary deformation of the surface will not be compatible with the manifestly 
conformally flat  form of the metric. For
example, any purely normal deformation of a non-planar surface will generally be inconsistent with
the representation; consistency will require a compensating tangential deformation to tag along.
Fortunately, this awkward technical point never needs to be addressed explicitly in the variational framework we introduce. 
It does, however,  manifest itself  in the conservation law for the stress tensor.
Whereas the tangential conservation laws would be expected to be satisfied identically in any
completely reparametrization invariant framework, where small tangential deformations of the
geometrical degrees of freedom are identified with reparametrizations (see \cite{Steigmannreparam}),
here they are not. This is also, of course, just as well: the Euler-Lagrange equations for the
spinor and the potential leave undetermined the off-diagonal component of the tangential stress; the
tangential components of the conservation law turn out to provide the differential equations
determining this missing component.

\vskip1pc\noindent
It is evident that the description of equilibrium in terms of a conserved stress tensor
possesses a number of interesting model-independent properties that are best appreciated by
extending the discussion to accommodate ``energy'' functionals of a more general kind. We will thus
examine geometrical energies  involving powers of the mean 
curvature.
The Canham-Helfrich model occurs as a special case.



\vskip1pc\noindent  
We begin in section \ref{sect2} with a brief review of relevant features of the spinor representation of  surfaces.
Our main results are contained in section \ref{sect3} where variational principles are developed within the spinor 
framework. In particular, the energy most relevant in condensed matter applications, consisting of a linear combination of 
area, the integrated mean curvature and bending energy is considered. 
We conclude with a brief discussion as well as a few suggestions for future work. Various useful identities are 
collected in a set of appendices.

\section{Dirac equation for surface spinors} \label{sect2}

In the generalized WE representation of the surface $\Sigma$ the three functions describing its embedding in
three-dimensional Euclidean space $\mathbb{E}^3$ are expressed in terms of a $2$-component spinor
field $\psi(z,\bar{z}) = \left(\psi_1(z,\bar{z}),\psi_2(z,\bar{z})\right)^T$ defined on a simply
connected domain $D$ of the complex plane $\mathbb{C}$ (the overbar denotes the complex conjugate).
This spinor is a  solution of the Dirac equation
\begin{equation} \label{diraceq}
\mathcal{D} \psi = 0\,,
\end{equation}
where $\mathcal{D}$ is the first order differential operator,
\begin{equation}\label{diracop}
\mathcal{D} = \left(
\begin{array}{cc}
0 & \partial_z \\
- \partial_{\bar{z}} & 0
\end{array} \right) +
\left(
\begin{array}{cc}
{\cal V} & 0 \\
0 & {\cal V}
\end{array} \right)\,,
\end{equation}
involving a real valued potential ${\cal V}$. This potential couples the spinor components; Eq.(\ref{diraceq}) reads
\begin{equation}
\partial_z \psi_2  = - {\cal V} \psi_1,  \quad
\partial_{\bar{z}} \psi_1  =  {\cal V} \psi_2\,.\\
\label{eqsDiraccomp}
\end{equation}
The spinor $\xi$ defined by $\xi=(-\bar{\psi}_2,\bar{\psi_1})^T$ also
satisfies the Dirac equation:
\begin{equation} \label{diraceqpsicc}
\mathcal{D} \xi = 0\,.
\end{equation}
We will also write $\xi = \zeta \bar{\psi}$, where
\begin{equation}
\zeta = \left (
\begin{array}{cc}
0 & -1\\
1 & 0
\end{array}
 \right)\,.
\end{equation}
Note also that $\xi$ unlike $\bar{\psi}$ transforms like $\psi$ under the action of $SU(2)$. A
linear combination of $\psi$ and $\zeta$ corresponds to a rotation of the surface, as described in
\ref{Approtsuf}.

\vskip1pc\noindent
The immersion of $D$ into $\mathbb{E}^3$ defined by $\psi(z,\bar{z})$ is given by
\begin{equation} \label{emdfunctgenWeier}
\mathbf{X}(z,\bar{z}) = \int_\gamma\, \bm{\phi} (w,\bar{w})\,,
\end{equation}
where $\bm{\phi}(z,\bar{z})$ is the vector-valued 
$1$-form with the following components\footnote{Here we interchange $\phi^1$ and $\phi^2$ with
respect to the more usual definitions in the literature, for example \cite{Taimanov}, 
as this facilitates the connection with the classical WE representation. The number of minus signs
appearing in the calculations is also reduced.}
\begin{subequations} \label{1formgenweier}
\begin{eqnarray}
\phi^1 (z,\bar{z}) & = & Re \left[(\bar{\psi}_2^{\phantom{2}2}-\psi_1^{\phantom{1}2} ) dz\right],\\
\phi^2 (z,\bar{z}) & = & Re \left[i ( \bar{\psi}_2^{\phantom{2}2} + \psi_1^{\phantom{1}2}) dz
\right], \\
\phi^3 (z,\bar{z}) & = & 2 Re \left[\psi_1 \bar{\psi}_2 dz \right]\,;
\end{eqnarray}
\end{subequations}
$\gamma$ is a path on the complex plane terminating at the point $z$. It is evident from their
definition that the embedding functions are invariant under complex conjugation, $\psi \to
\bar{\psi}$.

\vskip1pc\noindent It is simple to show that the $1$-forms (\ref{1formgenweier}) are closed in the
complex plane, i.e. $d \phi^i=0$: one uses the fact that the spinor $\psi$ satisfies the
Dirac equation, (\ref{diraceq}) and that ${\cal V}$ is real. Note that the
derivatives $\partial_{z} \psi_1$ and $\partial_{\bar z} \psi_2$ never need to be evaluated in this
argument; in addition, it should be remarked that the corresponding $1$-forms
constructed with the imaginary part in Eq.(\ref{1formgenweier}) are not closed.

\vskip1pc\noindent 
Closure and Stokes theorem together imply that the definition of the embedding functions
(\ref{emdfunctgenWeier}) is independent of the choice of the path $\gamma$ and thus well defined. It
is worth pointing out that the Dirac equation (\ref{diraceq}) appears to be the most general linear
equation for the spinor consistent with closure.

\subsection{Intrinsic geometry of the surface}

The tangent vectors to the surface adapted to this parametrization are $\textbf{e}_z = \partial_z
\textbf{X}$ and $\textbf{e}_{\bar{z}}= \Bar{\textbf{e}_z}$; in terms of the spinor
components $\textbf{e}_z$ is given by
\begin{equation} \label{weiertangvect}
\mathbf{e}_z = \frac{1}{2}\left(
\begin{array}{c}
\bar{\psi}_2^{\phantom{2}2}-\psi_1^{\phantom{1}2}\\
i \left( \bar{\psi}_2^{\phantom{2}2} + \psi_1^{\phantom{1}2} \right)  \\
2 \psi_1 \bar{\psi}_2
\end{array}
\right)\, .
\end{equation}
The vector-valued 1-form $\bm{\phi}$ defined in Eq.(\ref{1formgenweier}) can thus be expressed in
the alternative form  $\bm{\phi} = 2 Re \left( \textbf{e}_z dz \right).$

\vskip1pc \noindent
The two tangent vectors are null with respect to the  scalar product in $\mathbb{E}^3$ (denoted by
$\cdot$), in other words, their norms vanish: $\textbf{e}_z \cdot \textbf{e}_z = 0 =
\textbf{e}_{\bar{z}} \cdot \textbf{e}_{\bar{z}}$. However they are not orthogonal: $\textbf{e}_z
\cdot \textbf{e}_{\bar{z}} = 1/2 \, \lvert \psi \rvert^4$, where  $\lvert \psi \rvert^2 =
\psi^\dagger \psi = \lvert \psi_1\rvert^{2} + \lvert \psi_2\rvert^{2}$ and $\vert \psi_i \rvert^2 =
\bar{\psi}_i \psi_i$.

\vskip1pc\noindent
This parametrization is isothermal (or conformal): the line element is given by
\begin{equation}
ds^2 = \lvert \psi \rvert^4 \lvert d z \rvert^ 2;
\end{equation}
the induced metric on the surface  assumes the form
\begin{equation}
g_{ab} = \frac{\lvert \psi \rvert^4}{2} \left(
\begin{array}{cc}
0 & 1 \\
1 & 0
\end{array}
\right)\,.
\end{equation}
So the metric tensor is a Weyl rescaling of its Euclidean counterpart on the complex
plane, i.e., it is manifestly conformally flat. The fourth power of the spinor norm provides the conformal
factor. The intrinsic geometry of the surface is completely determined by $|\psi|$.

\vskip1pc\noindent
The residual reparametrization freedom consistent with the isothermal form of the metric is
captured by analytic functions. Under the holomorphic transformation $z\to w(z)$, $\bar z\to \bar w
(\bar z)$, one finds that
\begin{equation}
\psi_1(z,\bar{z})\to w'(z)^{1/2} \psi_1(w,\bar{w})\,,\quad \psi_2(z,\bar{z})\to \bar w'(\bar
z)^{1/2} \psi_2(w,\bar{w})\,, \quad {\cal V}(z,\bar{z}) \to |w'(z)|\, {\cal V}(w,\bar{w}) \,.
\label{trpsiV}
\end{equation}
These variables are thus scalar densities.

\vskip1pc\noindent
The determinant of the metric is $g = - 1/4 \lvert \psi \rvert^8 $, so that $\sqrt{g} = i/2 \lvert
\psi \rvert^4$ is imaginary. The area is given by
\begin{equation} \label{DWEArea}
A = \frac{i}{2} \int dz\wedge d\bar{z} \, |\psi|^4\,.
\end{equation}
Despite appearances, it is a real form: $\bar{d A} = -i/2 \lvert \psi \rvert^4 \, d \bar{z} \wedge dz
= d A$.

\subsection{Extrinsic geometry of the surface} \label{sectextgeom}

The normal vector to the surface $\Sigma$ defined by $\mathbf{n} = \mathbf{e}_z \times
\mathbf{e}_{\bar{z}}/ \sqrt{g} 
$, in terms of the spinor components is given by
\begin{equation} \label{DWEnormalvect}
\mathbf{n} = \frac{1}{\lvert \psi \rvert^2} \left(
\begin{array}{c}
2 \, Re \left( \psi_1 \psi_2 \right)\\
2 \, Im \left( \psi_1 \psi_2 \right)\\
\lvert \psi_1 \rvert^ 2 - \lvert \psi_2 \rvert^2
\end{array}
\right)\,.
\end{equation}

\vskip1pc \noindent The second fundamental form or extrinsic curvature tensor defined by $K_{ab} = -
\textbf{n} \cdot \partial_a \textbf{e}_b$, is given by
\begin{equation}
K_{ab}=
\left(
\begin{array}{cc}
{\cal A} & {\cal V} \lvert \psi \rvert^2\\
{\cal V} \lvert \psi \rvert^2 & \bar{{\cal A}}
\end{array}
\right)\,,
\end{equation}
where ${\cal A}$ denotes the Wronskian of $\psi_1$ and $\bar{\psi}_2$,
${\cal A}= \bar{\psi}_2 \partial_{z}\psi_1 - \psi_1 \partial_z \bar{\psi}_2$.
${\cal A}$ is invariant under rotation, as are the norm of $\psi$ and ${\cal V}$.
Under reparametrization, $z\to w(z)$, ${\cal A}$ transforms as a scalar density:
\begin{equation} \label{trA}
{\cal A}(z,\bar{z})\to  w'(z)^{2} {\cal A}\,.
\end{equation}
Besides the potential and the spinor norm,  the Wronskian is the other function of the spinor and its 
derivatives which occurs naturally in this framework. 
\vskip1pc\noindent

\vskip1pc \noindent 
The eigenvalues of the shape operator $K^a{}_b = g^{ac} K_{cb}$ ($g^{ab}$ is the inverse metric)
are the two principal curvatures, $C_1$ and $C_2$. The two symmetric curvature
invariants, (twice) the mean curvature $K=C_1+C_2$ and the Gaussian curvature $K_G=C_1C_2$ are given
respectively by
\begin{eqnarray}
K & = & \frac{4}{\lvert \psi \rvert^2} \, {\cal V}, \label{DWEK}\\
K_G & = & \frac{4}{\lvert \psi \rvert^8} \left({\cal V}^2 \lvert \psi \rvert^4 - \lvert {\cal A} \rvert^2
\right)\,.\label{DWEKGlg}
\end{eqnarray}
If ${\cal V} = 0$ the surface is minimal ($K=0$). By defining the components of the spinor $\psi$ in
terms of two analytic functions $f(z)$ and $g(z)$ as $\psi_1 = f^{1/2} g/\sqrt{2}$ and $\psi_2 =
\bar{f}^{1/2}/\sqrt{2}$ we obtain the $\bm{\phi}$ corresponding to the original WE representation
for minimal surfaces \cite{GrayFomenko} given by
\begin{equation}
\bm{\phi} = Re\left[ \frac{f}{2} \left(1-g^2,1+g^2,2 g\right)^T dz\right].
\end{equation}
The classical Gauss-Codazzi integrability condition on the surface geometry  (\ref{GaussWeingarten})
identifies
 (twice) the Gaussian curvature (\ref{DWEKGlg}) with the intrinsically defined Ricci scalar given by
(\ref{DWERicci}). 
Thus  $\lvert {\cal A} \rvert$ is completely determined once ${\cal V}$ and $|\psi|$ are known; it 
is a measure of the difference of the principal curvatures:
\begin{equation} \label{DWEModA}
\lvert {\cal A} \rvert^2 =
\frac{(C_1-C_2)^2}{16} \lvert \psi \rvert^8\,.
\end{equation}
The phase of ${\cal A}$ captures the principal directions; ${\cal A}$  vanishes only at umbilical points.

\subsection{Spinor presentation of bending energy} \label{secbendenergy}

Bending energy is quadratic in curvature. For a two-dimensional surface this
implies that it must also be scale invariant. There are several different possibilities. The
simplest of these is positive definite, proportional to 
\begin{equation}
H_1 = \frac{1}{2}\int dA \, K^{ab} K_{ab} = \frac{1}{2}\int dA \,
\left(C_1^{\phantom{1}2}+C_2^{\phantom{1}2}\right)\,,
\end{equation}
which vanishes on a planar region with $C_1=0=C_2$ ($K_{ab}=0$). The Canham-Helfrich energy
\cite{CanHel}
\begin{equation}
H_2= \frac{1}{ 2} \int dA \,  K^2= 4i \int dz\wedge d\bar{z}\, {\cal V}^2\,,
\end{equation}
which vanishes on a minimal surface, assumes a remarkably simple form in the spinor representation. The conformally invariant 
Willmore energy is given by the integrated squared difference of the principal curvatures
\cite{Willmore}
\begin{equation}
\label{eq:Willmore}
H_3 = \int dA \, \tilde K^{ab} \tilde K_{ab} = \frac{1}{2} \int dA \left(C_1-C_2\right)^2\,,
\end{equation}
where $\tilde K_{ab}= K_{ab}-\frac{1}{2} g_{ab} K$ is the trace-free curvature.
$H_3$ vanishes on a sphere. It is simple to see that
\begin{equation}
H_3  = 4 i \int dz\wedge d\bar{z} \frac{\lvert {\cal A} \rvert^2}{\lvert \psi \rvert^4} \,,
\end{equation}
involving the modulus of the Wronskian ${\cal A}$. 

\vskip1pc \noindent
Finally, one has the Gaussian bending energy, $H_4= \int dA\, K_G$ which is identified as the
Gauss-Bonnet topological invariant. It is clear that any quadratic bending energies can be
constructed using any two of $H_1,H_2,H_3$ and $H_4$. Choosing one of these as $H_4$, it is clear
that any one of the remaining three  is equivalent to any other modulo a boundary term associated with the Gauss-Bonnet invariant. In particular, the Willmore energy differs from the Canham-Helfrich bending 
energy by a topological term linear in the Gaussian curvature. Thus, despite the functional differences between the two, their Euler-Lagrange equations had better be identical. As a consistency check, we will confirm this fact  explicitly in the spinor framework in \ref{Compare}.

\section{The construction of the stress tensor} \label{sect3}

We are interested in examining the variational properties of various physically relevant energy
functionals of the surface geometry within the spinor framework. This may be the area,
bending energy or some other geometrical invariant. Such an invariant will generally be some scalar
density, $L$, constructed using the spinor and the real potential ${\cal V}$, integrated over the the 
surface \footnote{It will be convenient to work with a scalar density $L$, rather
than a scalar ${\cal L}$; the two are related by $L= \frac{1}{2}|\psi|^4 {\cal L}$.}
\begin{equation}
H = i \int dz \wedge d\bar{z} \, L(\psi_1,\bar{\psi}_1,\psi_2,\bar{\psi}_2,{\cal V})\,.
\end{equation}
Derivatives may also appear in $L$. Derivatives of $\psi$, if they occur, are expressible in terms
of $\psi$ itself and derivatives of $\ln \lvert \psi \rvert^2$, ${\cal V}$, and ${\cal A}$. Where
higher derivatives are involved it is not entirely straightforward in this framework to identify scalar
combinations of the variables appearing in the argument of $L$; it is facilitated, however, by
taking into account the transformation properties of the densities $\psi$, ${\cal V}$ given by
Eq.(\ref{trpsiV}) and of ${\cal A}$ given by Eq.(\ref{trA}), and forming products with weight zero.

\vskip1pc\noindent There are two sets of constraints that need to be accommodated when $H$ is varied with respect to these variables: the Dirac equation (\ref{diraceq}) implies that the variations in the spinor and the potential are not independent; we also must encode how to reconstruct the surface  in terms of the spinors. These constraints are enforced in the variational principle by introducing a number of Lagrange multipliers. Thus we construct the following functional:
\begin{align}
H_c = H &+ i \int dz \wedge d\bar{z} \, \lambda^{\dagger} \mathcal{D} \psi + i \int dz \wedge
d\bar{z} \, \lambda^T
\mathcal{D} \xi \nonumber \\
\phantom{=H} & + i \int dz \wedge d\bar{z} \, \mathbf{f}^z \cdot \left(\mathbf{e}_z - \partial_z
\mathbf{X}\right)+ i \int dz \wedge d\bar{z} \, \mathbf{f}^{\bar{z}} \cdot
\left(\mathbf{e}_{\bar{z}} - \partial_{\bar{z}} \mathbf{X}\right)\,,\label{eq:Hc}
\end{align}
where  $\mathcal{D}$ is defined by Eq.(\ref{diracop}) and 
$\mathbf{e}_z$ by Eq.(\ref{weiertangvect}).
The Lagrange multipliers $\lambda$ and $\mathbf{f}^z$ enforce respectively the Dirac equation
and the identification of the tangent vectors\footnote{As defined, both $\lambda$ and $\mathbf{f}^z$ are
densities.}. It is now legitimate to treat $\mathbf{X}$, $\psi$, $\bar{\psi}$ and ${\cal V}$ as
independent variables. The multiplier $\lambda$ is a $2$-component spinor defined by
$\lambda=\left(\lambda^1,\lambda^2\right)^T$
\,\footnote{The fact that that the Dirac equation (\ref{diraceqpsicc}) is the complex conjugate of
(\ref{diraceq}) is reflected in the fact that the multipliers enforcing these conditions are also
complex conjugates. In addition, $\textbf{f}^{\bar{z}}=\bar{\textbf{f}^z}$.}.
The multipliers $\mathbf{f}^z$ form a vector in $\mathbb{E}^3$ defined by
$\textbf{f}^z=\left(f^1,f^2,f^3\right)^T$. This construction is a spinor counterpart 
of that for parameterized surfaces \cite{auxil}. There are,
however, a number of important differences in its implementation.

\vskip1pc\noindent 
It may appear, at first, that the tangency constraints are redundant is the spinor framework. After all, any spinor satisfying the Dirac equation, will define a surface, and a naive counting of degrees of freedom suggests that nothing is amiss if they are dropped. This would certainly simplify the variational principle; unfortunately, it would also be wrong.
The multipliers $\mathbf{f}^z$, as we will see, will get identified with the stress tensor in the surface. Failure to enforce the constraint would leads one to the invalid conclusion that equilibrium solutions always have vanishing stress which is clearly not the case.

\vskip1pc \noindent
In components, the functional $H_c$ assumes the form\footnote{c.c. represents the complex conjugate expression.}
\begin{align}
H_c = H &+i \int dz \wedge d\bar{z} \, \bar{\lambda}^1 \left(\partial_z  \psi_2 + {\cal V}
\psi_1\right)+i \int dz \wedge d\bar{z} \, \bar{\lambda}^2 \left(\partial_{\bar{z}} \, \psi_1 - {\cal V}
\psi_2\right) + c.c. \nonumber \\
&+i \int dz \wedge d\bar{z} \left(\frac{f^1}{2}\left(\bar{\psi}_2^{\phantom{2}2}-\psi_1^{\phantom{1}2}\right) + i \frac{f^2}{2} \left( \bar{\psi}_2^{\phantom{2}2} + \psi_1^{\phantom{1}2} \right) +
f^3\psi_1 \bar{\psi}_2 - \mathbf{f}^z \cdot \partial_z \mathbf{X}\right) + c.c.\,.
\end{align}

\vskip1pc\noindent
\noindent Begin with the variation of the embedding functions $\textbf{X}$. Performing integration by parts to collect the derivatives of $\mathbf{f}^z$ and $\mathbf{f}^{\bar{z}}$ in a divergence, one finds
\begin{equation*}
\delta_\mathbf{X} H_c=i \int dz \wedge d\bar{z} \left( \partial_z \mathbf{f}^z \cdot \delta
\mathbf{X}-\partial_z (\mathbf{f}^z \cdot \delta \mathbf{X})\right)+c.c.\,.
\end{equation*}
Thus, the Euler-Lagrange derivative with respect to $\textbf{X}$, $\varepsilon_\textbf{X} \equiv
\delta H_c /\delta \textbf{X}$ is identified as
\begin{equation}
\varepsilon_\textbf{X} = \partial_z \mathbf{f}^z + \partial_{\bar{z}}
\mathbf{f}^{\bar{z}}\,. \label{DWEELX}\\
\end{equation}
Critical points of $H_c$ satisfy the Euler-Lagrange equations, $\varepsilon_\textbf{X}=0$. The solutions of the Euler-Lagrange equations are therefore described in terms of the conserved complex-valued surface ``stress tensor'', $\mathbf{f}^z$\,\footnote{With a densitized stress tensor  covariant differentiation is replaced by partial differentiation.}. It remains to construct $\mathbf{f}^z$ explicitly. This will involve the solution of the Euler-Lagrange equations for the variables $\psi$, $\bar{\psi}$ and ${\cal V}$.

\vskip1pc\noindent
Varying $H_C$ with respect to the potential ${\cal V}$ gives
\begin{equation*}
\delta_{\cal V} H_C = i \int dz \wedge d\bar{z} \left(\frac{\delta L}{\delta {\cal V}} +  \bar{\lambda}^1 \psi_1
+\lambda^1
\bar{\psi}_1 - \bar{\lambda}^2\psi_2 -\lambda^2\bar{\psi}_2\right) \delta {\cal V}\,;
\end{equation*}
the corresponding Euler-Lagrange equation is thus
\begin{equation}
\varepsilon_{\cal V} = \lambda^{\dagger} \sigma_3 \psi + \lambda^T \sigma_3 \bar{\psi}+ \frac{\delta
L}{\delta {\cal V}} = 0\,.\label{eq:varV}
\end{equation}
Here $\sigma_3$ is the Pauli matrix with $1$ and $-1$ along the diagonal. Since ${\cal V}$ is real, this equation is also real. The spinor $\lambda$ thus satisfies a single inhomogeneous linear algebraic equation in a two-dimensional complex vector space. One solution of this equation is given by
\begin{equation}
\lambda_i = -\frac{1}{2 \lvert \psi \rvert^2} \frac{\delta L}{\delta {\cal V}} \, \sigma_3
\psi\,.
\end{equation}
However, this solution is not unique. Let us write Eq.(\ref{eq:varV}) in the form $Re(\lambda^\dagger \sigma_3 \psi)=-\delta L/\delta {\cal V}$. Any spinor $\lambda$ with an imaginary projection onto $\sigma_3 \psi$ is evidently a solution of the homogeneous equation $Re(\lambda^\dagger \sigma_3 \psi)=0$. This equation has two solutions. The first of these is of the form
\begin{equation}
\lambda^r_h = i r \sigma_3 \psi,
\end{equation}
where $r$ is an arbitrary real-valued function. This is because
$\lambda_h^r{}^\dagger \sigma_3 \psi = - i |\psi|^2 r$ which is manifestly imaginary.
The second solution, which involves the spinor $\xi$, is given by
\begin{equation}
\lambda^c_h = \bar{c} \sigma_3 \xi,
\end{equation}
where $c$ is an arbitrary complex-valued function.
This solution satisfies $\lambda_h^\dagger \sigma_3 \psi =0$ on account of the orthogonality of
$\psi$ and $\xi$.
\vskip1pc\noindent Thus the complete solution $\lambda= \lambda_i+\lambda_h $ is given by
\begin{equation}
\lambda = \left(-\frac{1}{2 \lvert \psi \rvert^2} \frac{\delta L}{\delta {\cal V}} +i r \right)
\sigma_3 \psi\, + \bar{c} \sigma_3 \xi\,,
\end{equation}
the components of which read
\begin{equation} \label{DWElambdas}
\lambda^1 =  \left(-\frac{1}{2 \lvert \psi \rvert^2} \frac{\delta L}{\delta {\cal V}} +i r
\right) \psi_1 - \bar{c} \bar{\psi}_2, \qquad
\lambda^2 = \left(\frac{1}{2 \lvert \psi \rvert^2} \frac{\delta L}{\delta {\cal V}} - i r \right)
\psi_2\, - \bar{c} \bar{\psi}_1\,.
\end{equation}
This solution possesses 3 degrees of freedom per point: one for $r$ and two for $c$. To justify the counting, note that in terms of real-valued variables, the solution of equation (\ref{eq:varV}) describes a $3$-dimensional hyperplane in a $4$-dimensional vector space equipped with a (non-degenerate) inner product of signature $(+,+,-,-)$. It will be shown below that the ambiguity reflected in the functions $r$ and $c$ is a gauge artifact associated with the parametrization.

\vskip1pc\noindent
Let us introduce the quantities $f^+ = f^1 + i f^2$ and $f^- = f^1 - i f^2$. The Euler-Lagrange equations for the spinor $\psi$ are given by\footnote {Components of the complex conjugate of the spinor, $\bar{\psi}_1$ and $\bar{\psi}_2$, are varied independently of $\psi_1$ and $\psi_2$.}
\begin{subequations}
\begin{eqnarray}
\varepsilon_{\psi_1} &=&  - \psi_1 f^- + \bar{\psi}_2 f^3 + T^1= 0\,,
\label{weierELpsi1}\\
\varepsilon_{\psi_2} &=&  \psi_2 \bar{f^+} + \bar{\psi}_1 \bar{f}^3 + T^2= 0\,,
\label{weierELpsi2}
\end{eqnarray}
\end{subequations}
where
\begin{equation} \label{DWET1T2}
T^1 = \frac{\delta L}{\delta
\psi_1} -\partial_{\bar{z}}\bar{\lambda}^2 + \bar{\lambda}^1 {\cal V}\,, \qquad
T^2 = \frac{\delta L}{\delta \psi_2} - \partial_z \bar{\lambda}^1 -\bar{\lambda}^2 {\cal V}\,.
\end{equation}
together with their complex conjugate counterparts.

\vskip1pc\noindent Solving this set of equations for $\mathbf{f}^z$ is facilitated by by first expressing the Cartesian
components $f^+, f^-$ and $f^3$ in terms of their geometrically more relevant counterparts with
respect to the basis of tangent vectors adapted to the surface
$\{\textbf{e}_z,\textbf{e}_{\bar{z}},\textbf{n}\}$. The latter decomposition of $\textbf{f}^{z}$ is
given by
\begin{equation}
\sqrt{g} \, \mathbf{f}^z = i \left(\mathbf{f}^z \cdot \mathbf{e}_{\bar{z}}\,\mathbf{e}_z +
\mathbf{f}^z \cdot \mathbf{e}_z\, \mathbf{e}_{\bar{z}}\right) +\sqrt{g} \, \mathbf{f}^z \cdot
\mathbf{n}\, \mathbf{n}\,.
\end{equation}
The three projections can be expressed in terms of the Cartesian components, $f^+,f^-,f^3$, as follows
\begin{subequations}
\begin{eqnarray} \label{DWEfprojections}
f^{z}_{\phantom{z}z} & = & \mathbf{f}^z \cdot \mathbf{e}_z = \frac{1}{2} \left(\bar{\psi}_2^{\phantom{2}2} f^+ -
\psi_1^{\phantom{1}2} f^-  \right)+\psi_ 1
\bar{\psi}_2 f^3, \label{fzipezir}\\
f^{z}_{\phantom{z}\bar{z}} & = & \textbf{f}^z \cdot \textbf{e}_{\bar{z}} =
-\frac{1}{2} \left(\bar{\psi}_1^{\phantom{1}2}f^+ - \psi_2^{\phantom{2}2}
f^- \right)+ \bar{\psi}_1 \psi_2 f^3,
\label{fzipecczir}\\
f^{z} & = &\mathbf{f}^z \cdot \mathbf{n} = \frac{1}{\lvert \psi \rvert^2} \left(\bar{\psi}_1 \bar{\psi}_2 f^+ + \psi_ 1 \psi_2 f^- +(\lvert \psi_1 \rvert^2-\lvert \psi_2\rvert^2) f^3\right)\,; \label{fzipenir}
\end{eqnarray}
\end{subequations}
these three equations are now inverted in favor of $f^+$, $f^-$ and $f^3$ to obtain
\begin{subequations}
\begin{eqnarray}
f^+ & = & \frac{2}{\lvert \psi \rvert^4}\left(\psi_2^{\phantom{1}2} f^z_{\phantom{z}z} -
\psi_1^{\phantom{2}2} f^z_{\phantom{z}\bar{z}} + \lvert \psi \rvert^2 \psi_1 \psi_2 f^z\right),\\
f^- & = & \frac{2}{\lvert \psi \rvert^4}\left(-\bar{\psi}_1^{\phantom{1}2} f^z_{\phantom{z}z} +
\bar{\psi}_2^{\phantom{2}2} f^z_{\phantom{z}\bar{z}} + \lvert \psi \rvert^2 \bar{\psi}_1
\bar{\psi}_2 f^z\right),\\
f^3 & = & \frac{2}{\lvert \psi \rvert^4}\left(\bar{\psi}_1 \psi_2 f^z_{\phantom{z}z} + \psi_1
\bar{\psi}_2 f^z_{\phantom{z}\bar{z}}\right) +\frac{1}{\lvert \psi \rvert^2} \left(\lvert \psi_1 \rvert^2
- \lvert \psi_2 \rvert^2\right) f^z\,.
\end{eqnarray}
\end{subequations}
Substituting these expressions into the EL equations (\ref{weierELpsi1}) and 
(\ref{weierELpsi2}) one  obtains
\begin{eqnarray}
\varepsilon_{\psi_1} &=& \frac{2}{\lvert \psi \rvert^2} \bar{\psi}_1 f^{z}_{\phantom{z}z} -
\bar{\psi}_2 f^z + T^1= 0,\\
\varepsilon_{\bar{\psi}_2} & = & \frac{2}{\lvert \psi \rvert^2} \psi_2 f^{z}_{\phantom{z}z} +
\psi_1 f^z + \bar{T}^2=0\,.
\end{eqnarray}
In particular, the EL equations for $\psi_1$ and $\bar{\psi_2}$ (and the conjugates of these
equations) involve $f^{z}_{\phantom{z}z}$ and $f^{z}$ (and their complex conjugates) but not
$f^{z}_{\phantom{z}\bar{z}}$ (or its complex conjugate).
The tangential projection $f^{z}_{\phantom{z}\bar{z}}$ remains undetermined at this level.
At an algebraic level, this fact is related to the identity $\mathbf{e}_{z} = i \mathbf{e}_{z} \times \mathbf{n}$.

\vskip1pc \noindent
The combination $\psi_1 \, \varepsilon_{\psi_1}+\bar{\psi}_2 \, \varepsilon_{\bar{\psi}_2}$ determines $f^{z}_{\phantom{z}z}$:
\begin{equation}
\psi_1 \, \varepsilon_{\psi_1}+\bar{\psi}_2 \, \varepsilon_{\bar{\psi}_2} =  2
f^{z}_{\phantom{z}z}+\psi_1 T^1 +\bar{\psi}_2 \bar{T}^2=0\,.
\end{equation}
Using expressions (\ref{DWET1T2}) for $T^1$ and $T^2$ in this equation and solving
for $f^z_{\phantom{z}z}$ one finds
\begin{equation*}
f^z_{\phantom{z}z}  = -\frac{1}{2}\left(\psi_1 \frac{\delta L}{\delta \psi_1}+\bar{\psi}_2
\frac{\delta L}{\delta \bar{\psi}_2}\right) + \frac{1}{2} \partial_{\bar{z}}
\left(\lambda^1 \bar{\psi}_2 + \bar{\lambda}^2 \psi_1 \right)  - i {\cal V} \, Im
(\lambda^{\dagger}\psi)\,.
\end{equation*}
Now, by substituting into this last equation the expressions for $\lambda^1$ and $\lambda^2$ given
in (\ref{DWElambdas}) one finally obtains
\begin{equation} \label{DWEFZz}
f^z_{\phantom{z}\,z}  = -\frac{1}{2} \left(\psi_1 \frac{\delta L}{\delta
\psi_1}+\bar{\psi}_2 \frac{\delta L}{\delta \bar{\psi}_2}\right)+i \psi_1
\bar{\psi}_2 \partial_{\bar{z}} r -\frac{1}{2} \left(\bar{\psi}^{\phantom{2}2}_2 \partial_{\bar{z}} \bar{c} + \psi^{\phantom{2}2}_1 \partial_{\bar{z}} c \right)\,.
\end{equation}
Similarly $f^z$ is determined by the combination $\bar{\psi}_1
\, \varepsilon_{\bar{\psi}_2} - \psi_2 \, \varepsilon_{\psi_1}$:
\begin{eqnarray}
\bar{\psi}_1 \, \varepsilon_{\bar{\psi}_2} - \psi_2 \, \varepsilon_{\psi_1} & = & \lvert \psi \rvert^2
f^ z+\bar{\psi}_1 \bar{T}^2-\psi_2 T^1 = 0.
\end{eqnarray}
Substitution of expressions for $T^1$ and $T^2$ into this equation and solving for $f^z$ gives
\begin{eqnarray*}
f^z&=& \frac{1}{\lvert \psi \rvert^2}\left( \psi_2 \frac{\delta L}{\delta \psi_1} - \bar{\psi}_1 \frac{\delta L}{\delta \bar{\psi}_2} + \partial_{\bar{z}} \left(\lambda^1 \bar{\psi}_1 - \bar{\lambda}^2 \psi_2 \right)\right)\\
&+&\frac{1}{\lvert \psi \rvert^2}\left( \bar{\lambda}^2
\partial_{\bar{z}} \psi_2 - \lambda^1 \partial_{\bar{z}} \bar{\psi}_1 + {\cal V} (\bar{\lambda}^1 \psi_2+\lambda^2 \bar{\psi_1})\right)\,.
\end{eqnarray*}
Using once again  expressions  (\ref{DWElambdas}) in place of $\lambda^1$ and $\lambda^2$ along with the the expressions  (\ref{DWEdzpsi1}) and (\ref{DWEdczpsi2}) for the derivatives
of $\psi_1$ and $\psi_2$,  the following simplification results 
\begin{eqnarray} \label{DWEFZ}
f^z &=& - \partial_{\bar{z}} \left(\frac{1}{2 \lvert \psi \rvert^2}
\frac{\delta L}{\delta {\cal V}} \right) + \frac{1}{\lvert \psi \rvert^2 } \left( \psi_2 \frac{\delta L}{\delta \psi_1} - \bar{\psi}_1 \frac{\delta L}{\delta \bar{\psi}_2}\right)
\nonumber\\
&+& \frac{i}{\lvert \psi \rvert^2}(\lvert \psi_1 \rvert^2-\lvert \psi_2 \rvert^{2})
\partial_{\bar{z}} r+\frac{1}{\lvert \psi \rvert^2}\left(\psi_1 \psi_2 \partial_{\bar{z}}c - \bar{\psi}_1 \bar{\psi}_2 \partial_{\bar{z}} \bar{c} \right)\,.
\end{eqnarray}
The only ambiguity in the components $f^z_{\phantom{z}\,z} $ and $f^z $ of the stress tensor  is the one inherited from the solution of
Eq.(\ref{eq:varV}).

\vskip1pc \noindent
As noted previously, the Euler-Lagrange equations for the spinor and the potential leave completely undetermined the off-diagonal component $f^{z}_{\phantom{z}\bar{z}}$ of the tangential stress. At first, this appears to suggest that something is amiss. One must remember, however, that by representing the surface isothermally, one necessarily foregoes access to reparametrization invariance. This feature manifests itself in the tangential projections of the conservation law for the stress tensor. Whereas these equations would be satisfied identically in any completely reparametrization invariant framework, in this one they provide the differential equations determining the missing
component of the tangential stress.

\vskip1pc\noindent  Taking the projections of the conservation law $\varepsilon_\textbf{X}=0$,
where $\varepsilon_\textbf{X}$ is given by Eq.(\ref{DWEELX}), onto the tangent vectors provide the
equation
\begin{equation} \label{DWEELez}
\varepsilon_{\mathbf{X}} \cdot \mathbf{e}_z =  \partial_{\bar{z}}
f^{\bar{z}}_{\phantom{z}z}+|\psi|^4\partial_z \left(\frac{1}{|\psi|^4} f^z_{\phantom{z}z}\right) + 
{\cal A}\, f^z + \lvert \psi \rvert^2 {\cal V} f^{\bar{z}} = 0,
\end{equation}
along with its complex conjugate expression. Thus far, it has not been necessary to specify
explicitly the functional form of $L$. To solve Eq.(\ref{DWEELez}), we will 
suppose for simplicity that $L$ depends only on  $|\psi|^2$ and ${\cal V}$, undifferentiated. These differential equations can then be solved for the missing component of the stress tensor. The most general solution is given by
\begin{equation} \label{DWEFZzb}
f^{z}_{\phantom{z}\bar{z}} = \frac{\bar{{\cal A}}}{2 \, \lvert \psi \rvert^2} \frac{\delta L}
{\delta {\cal V}}+ i \bar{\psi}_1 \psi_2 \partial_{\bar{z}} r +\frac{1}{2} \left(\psi^{\phantom{2}2}_2 \partial_{\bar{z}}c +\bar{\psi}^{\phantom{2}2}_1 \partial_{\bar{z}}\bar{c}\right)+\bar{h}(\bar{z})\,,
\end{equation}
where $h(z)$ is an arbitrary function.

\vskip1pc\noindent
The projection onto the normal vector provides the ``shape'' equation
\begin{equation}
\varepsilon_{\mathbf{X}} \cdot \mathbf{n} = \partial_z f^z + \partial_{\bar{z}} f^{\bar{z}}-\frac{K}{2} \left(f^z_{\phantom{z}z}+f^{\bar{z}}_{\phantom{z}\bar{z}}\right) - \frac{2}{\lvert \psi \rvert^4}
\left({\cal A} f^z_{\phantom{z}\bar{z}}+\bar{{\cal A}} f^{\bar{z}}_{\phantom{z}z}\right)=0\,.
\label{DWEELn}
\end{equation}

\vskip1pc\noindent
We are now in a position to examine the different ambiguities which have arisen in our construction of the stress tensor. The first of these originates in the solution of the Euler-Lagrange equation for the potential.
\vskip1pc\noindent
The contribution to the stress originating in the homogeneous solution $\lambda^c_h$ is given by
\begin{eqnarray}
\lvert \psi \rvert^4 \mathbf{f}_c^z &= &\partial_{\bar{z}} c \left(\psi^{\phantom{1}^2}_2\mathbf{e}_{z}-\psi^{\phantom{1}^2}_1
\mathbf{e}_{\bar{z}}+\lvert \psi \rvert^2 \psi_1 \psi_2 \mathbf{n}\right) \nonumber \\
&+& \partial_{\bar{z}} \bar{c} \left( \bar{\psi}^{\phantom{1}^2}_1 \mathbf{e}_{z}-\bar{\psi}^{\phantom{1}^2}_2
\mathbf{e}_{\bar{z}}-\lvert \psi \rvert^2\bar{\psi}_1 \bar{\psi}_2
\mathbf{n}\right) \,.
\end{eqnarray}
Decomposing the complex-valued function $c$ into its real and imaginary parts, $c=c_x + i c_y$, it
is possible to express this contribution as the partial derivative of a space vector
\begin{equation}
\mathbf{f}_c^z = i \partial_{\bar{z}} \left( c_y \hat{\mathbf{x}}_1 + c_x \hat{\mathbf{x}}_2\right)\,,
\end{equation}
where the fact that the basis vectors of $\mathbb{E}^3$ are constant has been used (holomorphic will
do). With respect to the surface adopted basis these three vectors are given by
\begin{subequations}
 \begin{eqnarray}
\hat{\mathbf{x}}_1 &=& \frac{1}{\lvert \psi \rvert^4} \left( (\psi_2^{\phantom{2}2} -
\bar{\psi}_1^{\phantom{1}2})\mathbf{e}_z+(\bar{\psi}_2^{\phantom{2}2} -
\psi_1^{\phantom{1}2})\mathbf{e}_{\bar{z}}+\lvert \psi \rvert^2 (\psi_1 \psi_2+\bar{\psi}_1
\bar{\psi}_2)\mathbf{n}\right),\\
\hat{\mathbf{x}}_2 &= & \frac{i}{\lvert \psi \rvert^4} \left( -(\psi_2^{\phantom{2}2} +
\bar{\psi}_1^{\phantom{1}2})\mathbf{e}_z+(\bar{\psi}_2^{\phantom{2}2} +
\psi_1^{\phantom{1}2})\mathbf{e}_{\bar{z}}-\lvert \psi \rvert^2 (\psi_1 \psi_2-\bar{\psi}_1
\bar{\psi}_2)\mathbf{n}\right),\\
\hat{\mathbf{x}}_3 & = & \frac{2}{\lvert \psi \rvert^4} \left(\bar{\psi}_1 \psi_2 \mathbf{e}_z +
\psi_1 \bar{\psi}_2 \mathbf{e}_{\bar{z}} + \frac{1}{2} \lvert \psi \rvert^2(\lvert \psi_1
\rvert^2-\lvert \psi_2 \rvert^2)\mathbf{n}\right)\,.
\end{eqnarray}
\end{subequations}
Similarly, the contribution to the stress tensor originating in $\lambda^r_h$ is given by
\begin{equation}
\mathbf{f}^z_r = i \partial_{\bar{z}} \left( r \hat{\mathbf{x}}_3 \right)\,,
\end{equation}
so that in full the contribution to the stress tensor arising from the homogeneous solution can be expressed as the partial derivative of a real-valued space vector $\textbf{V}$,
\begin{equation}
\mathbf{f}^z_h = i \partial_{\bar{z}} \mathbf{V}\,,
\end{equation}
where $\textbf{V}$ is given by $\mathbf{V} = \left(c_y \hat{\mathbf{x}}_1 + c_x \hat{\mathbf{x}}_2+r \hat{\mathbf{x}}_3\right)$. Since $\mathbf{V}$  is real-valued, $\textbf{f}_h^z$ has zero divergence: $\partial_z \textbf{f}_h^z+\partial_{\bar{z}} \textbf{f}_h^{\bar{z}}=0$. It is a null tensor which is automatically conserved and
 does not contribute to the shape equation (\ref{DWEELn}). 
Therefore, as claimed above, it is legitimate to neglect this contribution to the stress tensor and retain only the part arising from the inhomogeneous solution $\lambda_i$.

\vskip1pc\noindent
Unlike the canonical stress tensor in the parametrized description of a surface, its counterpart 
in this framework  is not unique. 

\vskip1pc \noindent
The ambiguity associated with the arbitrary function $h(z)$ appearing in the solution of 
Eq.(\ref{DWEELez}) given by Eq.(\ref{DWEFZzb}) appears to be of a more serious nature.  It is not simply a
gauge artifact, contributing as it does to the shape equation a term
\begin{equation*}
\approx\frac{2}{\lvert \psi \rvert^4}
\left({\cal A} \bar{h}(\bar{z})+\bar{{\cal A}}
h(z)\right)\,.
\end{equation*}
However, the normal projection of the Euler-Lagrange derivative $\varepsilon_{\mathbf{X}} \cdot \mathbf{n}$ associated with any reparametrization invariant energy function had better form a scalar (density). Under a holomorphic transformation $ z \rightarrow w(z)$ the factor ${\cal A}/ \lvert \psi \rvert^4$ transforms as 
\begin{equation}
\frac{{\cal A}(z,\bar{z})}{\lvert \psi (z,\bar{z}) \rvert^4} \rightarrow \frac{w'(z)}{\bar{w}'(\bar{z})} \frac{{\cal A}(w,\bar{w})}{\lvert \psi (w,\bar{w})\rvert^4}\,.
\end{equation}
In order to form a scalar $h(z)$ should transform as $h(z) \rightarrow w'(z)/\bar{w}'(\bar{z}) h(w)$,  which is  
 not a holomorphic function.
Consistency 
requires that $h$  must vanish.

\vskip1pc \noindent  In summary, for a functional which depends on $|\psi|$ and ${\cal V}$, the components of the stress tensor assume the simple form 
\begin{subequations} \label{DWESTfnl}
\begin{eqnarray}
f^z_{\phantom{z}z}  &=& -\frac{1}{2} \left(\psi_1 \,\frac{\partial L}{\partial
\psi_1}+\bar{\psi}_2 \,  \frac{\partial L}{\partial \bar{\psi}_2}\right),\\
f^{z}_{\phantom{z}\bar{z}} &=& \frac{\bar{{\cal A}}}{2 \, \lvert \psi \rvert^2} \frac{\partial L}
{\partial {\cal V}}\,,\\
f^z & = & -\partial_{\bar{z}} \left(\frac{1}{2 \lvert \psi \rvert^2}
\frac{\partial L}{\partial {\cal V}} \right) \,.
\end{eqnarray}
\end{subequations}
The Euclidean invariance of the energy and the identification of $\mathbf{f}^z$ with the stress
tensor within this framework is discussed in  \ref{strtrqtnsr}. A more general dependence on the 
spinor and potential is considered in \ref{Compare}.

\subsection{Canham-Helfrich energy}

In soft matter applications, the surface energy will typically be a sum of bending energy,
a term linear in $K$ reflecting an asymmetry between the two sides of the surface, and an area term
associated with a constraint or penalty on the total area \cite{Seifert},
\begin{equation}
L = {1\over 2} \kappa L_2 + \beta L_1 + \sigma L_0\,,
\end{equation}
where $L_n =  1/2 |\psi|^4 K^n$. 
 
\vskip1pc\noindent One finds that
\begin{equation}
\frac{\partial L_n}{\partial{\psi_i}} =  (1-\frac{n}{2})
\lvert \psi \rvert^2 K^n \bar{\psi}_i \qquad \mbox{and} \qquad
\frac{\partial L_n}{\partial {\cal V}} = 2 \,  \,n \,\lvert \psi \rvert^2 K^{n-1}\,.
\end{equation}
From equations (\ref{DWESTfnl})
one reads off the contribution of $L_n$ to the various
components of
the stress tensor:
\begin{subequations}
\begin{eqnarray}
f_n{}^z_{\phantom{z}z} &=& \frac{(n-2)}{4} |\psi|^4 K^n\,, \\
f_n{}^{z}_{\phantom{z}\bar{z}} & = & \, n \, \bar{{\cal A}} K^{n-1}\,,\\
f_n{}^z & = & - \, n \, \partial_{\bar{z}} K^{n-1}\,.
\end{eqnarray}
\end{subequations}
It is simple to confirm that these expressions reproduce the well-known result \cite{Stress},
\begin{equation}
\mathbf{f}_n^a=  K^{n-1}( n K^{ab} - K g^{ab}) \,\mathbf {e}_b - n \nabla^a K^{n-1}\, \mathbf{n}\,,
\label{stressn}
\end{equation}
in this particular parametrization.
The corresponding contribution to the normal component of the Euler-Lagrange derivative (\ref{DWEELn}) is also easily shown to be given by
\begin{equation}
{\cal E}_n = - n  \Delta K^{n-1} + K^{n-1} \left(2 n K_G  - (n-1) K^2 \right) \,.
\end{equation}

\subsubsection*{n=0: Area}
For surface area, corresponding to $L_0 = 1/2 |\psi|^4$,
only the diagonal tangential stress is
non-vanishing:
\begin{equation}
f^z_{\phantom{z}z} = - \frac{1}{2}|\psi|^4, \quad f^{z}_{\phantom{z}\bar{z}} = 0, \quad f^z = 0\,.
\end{equation}
The Euler-Lagrange derivative is given by ${\cal E}_0=K$; the critical points of area  are minimal surfaces satisfying $K=0$ or ${\cal V}=0$.

\subsubsection*{n=1: Integrated mean curvature}

For an energy proportional to the mean curvature, corresponding to $L_1 = 1/2 |\psi|^4 K$ ,
the components of the tangential stress tensor are
\begin{equation}
f^z_{\phantom{z}z} = -\frac{1}{4} |\psi|^4 K, \quad f^{z}_{\phantom{z}\bar{z}} = \bar{{\cal A}}\,,
\end{equation}
and the normal stress vanishes, $f^z = 0$.
The Euler-Lagrange derivative  ${\cal E}_1= \mathcal{R}$; 
the critical points are developable with vanishing Gauss curvature.

\subsubsection*{n=2: Canham-Helfrich bending energy $L_B=L_2/2= 4 {\cal V}^2$}

The components of the stress tensor are
\begin{equation} \label{DWEfcmpWlmr}
f^z_{\phantom{z}z} = 0, \quad f^{z}_{\phantom{z}\bar{z}} = \bar{{\cal A}} K, \quad f^z =
-\partial_{\bar{z}} K\,,
\end{equation}
and the Euler-Lagrange derivative given by 
\begin{equation}
{\cal E}_B= -\Delta K +K \left(2 K_G -\frac{1}{2}K^2\right)\,.
\end{equation}
The vanishing of $f^z_{\phantom{z}z}$ in this case is a manifestation of scale invariance.

\vskip1pc\noindent
It was pointed out in the paragraph following Eq.(\ref{eq:Hc}) that if the constraints--defining
the stress tensor--relating the tangents to the embedding variables are not enforced,
the Euler-Lagrange equations obtained are incorrect. To underscore this point, consider  the
consequence of dropping this constraint so that ${\mathbf f}^a=0$; the surface states are
stress-free. In the case of pure bending, setting ${\mathbf f}^a=0$ in Eq.(\ref{stressn}) for $n=2$
implies that $K=0$ or $K_{ab}=g_{ab} K/2$; the latter possibility implies a spherical geometry. The
only stress-free solutions are thus minimal surfaces or spheres.  Even though the variations of
$\psi$ and ${\cal V}$ are consistent with the Dirac equation, and thus represent a surface,
without the additional constraints, the corresponding Euler-Lagrange equations do not describe the
critical points of the surface problem correctly.

\section{Discussion}

We have provided a variational framework, tailored to the representation of the
surface geometry in terms of a spinor field interacting through a potential, to describe the equilibrium properties of this surface. The construction of a conserved stress tensor lies at the center of this framework.    
Our primary aim, of course, was not to provide yet another construction of the stress tensor. Even
if this derivation does provide significant new insight into its relationship with the surface
geometry, there are other derivations; rather it has been to examine how these variables can be
reconciled with  surface variational principles.  One is now in a position to import the techniques of complex analysis 
to re-examine various inherently non-linear problems in soft matter where our toolbox has been found wanting. 

\vskip1pc\noindent 
A problem that lends itself to be treated using the spinor representation of the surface
geometry is a very old one  that has been the focus of a revival of interest in recent years: how
does a surface bend when its local geometry is constrained; a good example is provided 
by an unstretchable planar sheet of paper. Contrary to initial expectations, this is not a simple problem
\cite{benpom,cerdamaha,Witten}. The general configuration will consist of piecewise developable
surfaces (with vanishing Gaussian curvature) meeting along a set of ridges.  
Such surfaces assume a particularly simple form in the spinor representation on account of Eq.(\ref
{DWERicci}). If the Gaussian curvature vanishes, $|\psi|$ itself is harmonic. 
Unfortunately, developable surfaces do not generally occur as solutions of the unadorned
Euler-Lagrange equations for bending: the correct Euler-Lagrange equations possess an additional
term arising from the local constraint on the metric that needs to be imposed to maintain flatness
under variation;  they do not minimize bending energy unless one imposes this  additional local 
constraint \cite{Paperfold}. One also finds that the boundary conditions that are appropriate will
reflect the constraint on the metric. A simple but non-trivial example is provided by a cone; it
does not minimize bending energy unless the metric is constrained to be flat. This is why paper
folds into cones but fluid membranes do not! This local constraint is very different in nature from
the structural constraints appearing in the auxiliary framework; whereas the role of the latter set
of constraints is to ensure that the spinors and the potential are consistent with some spatial
geometry, they do not place any constraint on the geometry itself. It is remarkable therefore that
the local constraint can be treated technically on an identical footing \cite{Paperfold}. 
The extra term takes the form of a linear coupling of the corresponding tangential stress to
extrinsic curvature. 
We are currently examining this problem within the  spinor framework.

\vskip1pc\noindent 
There are several interesting directions for future work. 
Statistical field theory is an obvious one, where we have been largely limited to Gaussian approximations
using traditional methods.   There is, however,  still some spadework to be done 
before one is in a position to do this with any confidence.  It would be useful to first address 
a few questions of an elementary nature: how does one describe perturbation theory about
some given equilibrium surface in this representation? Does the simple form of the bending energy
have a counterpart in the expansion at second order? In this context, it is clear that even if one
is interested in minimal surfaces described by the WE representation with ${\cal V}=0$, one needs to
introduce a potential in the variation.  Finally, it would appear that 
 the coupling of electrons to the curved surface geometry in graphene 
 calls naturally for 
a spinor description of the surface.



\vskip1pc\noindent  Partial support from DGAPA PAPIIT grant IN114510 is acknowledged.

\begin{appendix}

\setcounter{equation}{0}
\renewcommand{\thesection}{Appendix \Alph{section}}
\renewcommand{\thesubsection}{A. \arabic{subsection}}
\renewcommand{\theequation}{A.\arabic{equation}}

\section{Rotation of the surface in $\mathbb{E}^3$} \label{Approtsuf}

\noindent The Dirac equation is not $SU(2)$ invariant. Nor are rotations of the surface realized
in what would appear to be the ``obvious'' way by the action of $SU(2)$ on the spinor $\psi$ using
the local isomorphism $SO(3) = SU(2)$. Note, however, that the surface which corresponds to the
spinor $\xi$ is the same as that obtained from $\psi$ under a rotation by an angle $\pi$ around the
$\hat{\textbf{x}}_2$ axis, i.e. under the substitution $\psi \rightarrow \xi$ the components of
$\textbf{X}$ change as $X^1 \rightarrow -X^1, X^2 \rightarrow X^2, X^3 \rightarrow -X^3$. More
generally, the transformed spinor $\tilde{\psi}=\bar{a} \psi\ + b\, \xi$, where $a$ and $b$ are
two complex numbers, also satisfies the Dirac equation so that it, as well, describes a surface
$\tilde{\textbf{X}}$. If $|a|^2 + |b|^2 =1$, $\tilde{\psi}$ is indirectly related to $\psi$ by an element $U$
of $SU(2)$ given by
\begin{equation}
U = \left(
\begin{array}{cc}
a&-\bar{b}\\
b&\bar{a}
\end{array}
\right)\,,
\end{equation}
which can also be represented as $U = e^{- i \frac{\vartheta}{2} \bm{\sigma}\cdot
\hat{\mathbf{r}}}$, where $\hat{\mathbf{r}}$ is a unit vector in $\mathbb{E}^3$ and $\bm{\sigma}$ is
the vector in $\mathbb{E}^3$ with the Pauli matrices as its components. Because of the local
isomorphism $SO(3) = SU(2)$ this identifies the element $\sf{R}$ of $SO(3)$ given by ${\sf R} = e^{i
\vartheta \mathbf{J} \cdot \hat{\mathbf{r}}}$, where $\textbf{J}$ is a vector in $\mathbb{E}^3$ with
the infinitesimal generators of rotations
as its components, so that ${\sf R}$ describes a anticlockwise rotation by an angle $\vartheta$
around the direction $\hat{\mathbf{r}}$.  Evidently $\tilde{\psi} \neq U \psi$, thus the manner in
which $U$ relates $\tilde{\psi}$ and $\psi$ is not through the usual action of $SU(2)$, but with its
associated rotation $R$ that relates the embedding functions of both spinors, namely the embedding
functions $\tilde{\mathbf{X}}$ describe a rotation of $\mathbf{X}$, i.e. $\tilde{\mathbf{X}}= {\sf
R} \mathbf{X}$ \cite{Taimanov}. If $|a|^2 + |b|^2 \ne 1$, the transformation $\psi \to \bar{a} \psi\
+ b\, \xi$ describes a rotation accompanied by a scaling.

\setcounter{equation}{0}
\renewcommand{\thesection}{Appendix \Alph{section}}
\renewcommand{\thesubsection}{B. \arabic{subsection}}
\renewcommand{\theequation}{B.\arabic{equation}}

\section{The Laplace-Beltrami operator and the Ricci scalar}

The Laplace-Beltrami operator $\Delta$ on the complex plane is given by
\begin{equation}
\Delta =  \frac{4}{\lvert \psi \rvert^4} \partial_z \partial_{\bar{z}}\,.
\end{equation}
The non-vanishing Christoffel symbols constructed with the metric $g_{ab}$ are
\begin{equation}
\Gamma^{z}_{\phantom{z}zz} = g^{z \bar{z}} \partial_z g_{z \bar{z}}
= \partial_z \ln g_{z\bar{z}}=2 \,\partial_z \ln \lvert \psi \rvert^2\,,
\end{equation}
and its complex conjugate $\Gamma^{\bar{z}}_{\phantom{z}\bar{z}\bar{z}}$.
Likewise, the only non-vanishing components of the Riemann tensor in the conformal parametrization are
\begin{equation}
R^{z}_{\phantom{z}z\bar{z}z}=\frac{\lvert \psi \rvert^4}{2} \Delta \ln \lvert \psi \rvert^2 =
R^{\bar{z}}_{\phantom{z} \bar{z}z\bar{z}}\,.
\end{equation}
The Ricci tensor (proportional to the metric) and the Ricci scalar are
\begin{equation} \label{DWERicci}
R_{z\bar{z}} = - \frac{\lvert \psi \rvert^4}{2} \Delta \ln \lvert \psi \rvert^2, \qquad \mathcal{R}
= - 2 \Delta \ln \lvert \psi \rvert^2\,.
\end{equation}

\setcounter{equation}{0}
\renewcommand{\thesection}{Appendix \Alph{section}}
\renewcommand{\thesubsection}{C. \arabic{subsection}}
\renewcommand{\theequation}{C.\arabic{equation}}

\section{Identities for $\partial_z \psi_1$ and $\partial_z \bar{\psi}_2$}
 
The two symmetric curvature scalars depend on $\psi$ only through the combinations $|\psi|^2$ and
${\cal A}$. It is useful to possess identities for the missing partial derivatives,  $\partial_z
\psi_1$ and $\partial_z \bar{\psi}_2$, in terms of these variables. 

\vskip1pc \noindent
Begin by differentiating $\lvert \psi \rvert^2$ with respect to $z$, using the
fact that $\psi$ satisfies the Dirac equation, to obtain
\begin{equation}
\label{pzealpha}
\partial_z \lvert \psi \rvert^2 = \psi_2 \partial_z \bar{\psi}_2 + \bar{\psi}_1 \partial_z
\psi_1\,.
\end{equation}
Now multiply across by $\psi_1$ and add $\psi_2 {\cal A}$ to both sides to get
\begin{equation*}
\psi_2 {\cal A} + \psi_1 \partial_z \lvert \psi \rvert^2 = \lvert \psi \rvert^2 \partial_z
\psi_1\,,
\end{equation*}
we thus obtain the identity
\begin{equation} \label{DWEdzpsi1}
\partial_z \psi_1 = \psi_1 \partial_z \ln \lvert \psi \rvert^2 + \frac{1}{\lvert \psi \rvert^2}
\psi_2 {\cal A}\,.
\end{equation}
Multiplying Eq.(\ref{pzealpha}) across by $\bar{\psi}_2$ and subtracting $\bar{\psi}_1 {\cal A}$ from both sides yields
\begin{equation*}
\bar{\psi}_2 \partial_z \lvert \psi \rvert^2 - \bar{\psi_1} {\cal A} = \lvert \psi \rvert^2
\partial_z \bar{\psi}_2\,,
\end{equation*}
giving the second identity
\begin{equation} \label{DWEdczpsi2}
\partial_z \bar{\psi}_2 = \bar{\psi}_2 \partial_z \ln \lvert \psi \rvert^2 - \frac{1}{\lvert \psi
\rvert^2} \bar{\psi}_1 {\cal A}\,.
\end{equation}
All derivatives of $\psi$ can now be expressed in the compact form,
\begin{equation}
\partial_z \psi = \left(
\begin{array}{cc}
\partial_z  \ln \lvert \psi \rvert^2 & \lvert \psi \rvert^{-2} {\cal A}\\
-{\cal V} & 0
\end{array}
\right) \psi\,, \qquad
\partial_{\bar{z}} \psi = \left(
\begin{array}{cc}
0 & {\cal V}\\
-\lvert \psi \rvert^{-2} \bar{{\cal A}}& \partial_{\bar{z}} \ln \lvert \psi \rvert^2
\end{array}
\right) \psi\,.
\end{equation}

\setcounter{equation}{0}
\renewcommand{\thesection}{Appendix \Alph{section}}
\renewcommand{\thesubsection}{D. \arabic{subsection}}
\renewcommand{\theequation}{D.\arabic{equation}}

\section{Gauss-Weingarten equations and the integrability conditions} \label{GaussWeingarten}

The Gauss-Weingarten equations, $\partial_a \mathbf{e}_a = \Gamma^c_{\phantom{c}ab} \mathbf{e}_c -
K_{ab} \mathbf{n}$ and $\partial_a \mathbf{n}= K_a{}^b\, \mathbf{e}_b$, describe how the adopted
frame $\{\mathbf{e}_a, \mathbf{n}\}$ changes as it is moved across the surface. In the generalized
WE representation, by writing the adapted basis to the surface as $\textbf{E} =
\{\textbf{e}_z,\textbf{e}_{\bar{z}},\textbf{n}\}$, these equations  can be expressed in the compact
form \cite{Bobenko}
\begin{equation}
\partial_z \textbf{E} = {\sf M}\,\mathbf{E} \qquad \partial_{\bar{z}} \mathbf{E} = {\sf N}\,\mathbf{E},
\end{equation}
where the linear transformations ${\sf M}$ and ${\sf N}$ are given by
\begin{equation}
{\sf M}=\left(
\begin{array}{ccc}
2 \partial_z \ln \lvert \psi \rvert^2 & 0 & - {\cal A}\\
0 & 0 &-\frac{1}{4} K \lvert \psi \rvert^4\\
\frac{1}{2} K & \frac{2}{\lvert \psi \rvert^4} {\cal A} & 0
\end{array}
\right), \quad
{\sf N}=\left(
\begin{array}{ccc}
0 & 0 & -\frac{1}{4} K \lvert \psi \rvert^4\\
0 & 2 \partial_{\bar{z}} \ln \lvert \psi \rvert^2 & -\bar{{\cal A}}\\
\frac{2}{\lvert \psi \rvert^4} \bar{{\cal A}}& \frac{1}{2} K & 0
\end{array}
\right)\,.
\end{equation}

\vskip1pc \noindent 
The integrability condition $\textbf{E}_{z\bar{z}}=\textbf{E}_{\bar{z}z}$ leads to
$ {\sf M}_{\bar{z}} - {\sf N}_z +[{\sf M},{\sf N}] = 0$. From this condition and the linear
independence of basis $\textbf{E}$ one obtains the following
relations
\begin{subequations}
\begin{eqnarray}
\partial_z \partial_{\bar{z}} \ln \lvert \psi \rvert^2 & = & \frac{\lvert
{\cal A} \rvert^2}{\lvert \psi \rvert^4} - {\cal V}^2 = \frac{\lvert {\cal A} \rvert^2}{\lvert \psi \rvert^4} -\frac{\lvert \psi \rvert^4}{16} K^2 \,,\label{DWEGC}\\
\partial_{\bar{z}} {\cal A} & = & \lvert \psi \rvert^2 \partial_z {\cal V} - {\cal V} \partial_z \lvert \psi \rvert^2 = \frac{\lvert \psi \rvert^4}{4} \partial_z K\, ,\label{DWECM}
\end{eqnarray}
\end{subequations}
which are fulfilled by spinors and potential satisfying the Dirac-type Eq.(\ref{diraceq}). So it is
unnecessary to implement them in the effective functional when performing the variational
principle.\\
Equation (\ref{DWEGC}) is the Gauss-Codazzi equation since it is equivalent to the equation
\begin{equation}
\mathcal{R} = K^2-K_{ab}K^{ab} = K^2 - (K_{zz}K^{zz}
+2K_{z\bar{z}}K^{z\bar{z}}+K_{\bar{z}\bar{z}}K^{\bar{z}\bar{z}}).
\end{equation}
Likewise, equation (\ref{DWECM}) and its complex conjugate are the Codazzi-Mainardi equations for they are  equivalent to the
pair of equations
\begin{equation}
\nabla_{a} K_{bc} = \nabla_{b} K_{ac} \qquad \text{or}  \qquad \nabla_{\bar{z}} K_{zz} = \nabla_{z}
K_{\bar{z}z}\quad \text{and its c.c.}\,.
\end{equation}

\setcounter{equation}{0}
\renewcommand{\thesection}{Appendix \Alph{section}}
\renewcommand{\thesubsection}{E. \arabic{subsection}}
\renewcommand{\theequation}{E.\arabic{equation}}

\section{Conserved stress and torque} \label{strtrqtnsr}

The current associated with the variation of the embedding functions, denoted by $Q_\textbf{X}$, is identified with the boundary term in the variation of the functional, $H_c$, with respect to $\mathbf{X}$, or
\begin{equation}
Q_\textbf{X} = -i \int dz \wedge d\bar{z} \left( \partial_z (\mathbf{f}^z \cdot \delta
\mathbf{X})+\partial_{\bar{z}} (\mathbf{f}^{\bar{z}} \cdot \delta
\mathbf{X})\right)\,.
\end{equation}
The current associated with the variations in the spinor field is given by
\begin{equation}
Q_{\psi_1} = i \int dz \wedge d\bar{z} \partial_{\bar{z}}(\bar{\lambda}^2
\delta \psi_1), \qquad Q_{\psi_2} = i \int dz \wedge d\bar{z}
\partial_z(\bar{\lambda}^1 \delta \psi_2),
\end{equation}
together with their complex conjugate expressions.

\vskip1pc \noindent
The potential does not give rise to a current. Thus the complete variation of
$H_c$ is given by
\begin{equation}
\delta H_c = i \int dz \wedge d\bar{z} \, \bm{\varepsilon} \cdot \delta \mathbf{X} + Q\,,
\end{equation}
where the total current $Q$ is given by $Q = Q_{\textbf{X}}+\left(Q_{\psi_1}+Q_{\psi_2}+c.c.\right)$. When the Euler-Lagrange equations are satisfied, $\delta H_c= Q$.

\vskip1pc \noindent
Consider now a patch of surface bounded by a set of closed curves. To determine the change in the equilibrium energy under the translation of one of these curves (say $\Gamma$), consider a deformation $\delta {\bf X}$ that reduces to a translation $\delta {\bf a}$ on this curve and vanishes on the remaining boundaries. The spinor field transforms trivially under translation. Making use of the Gauss theorem the change in the energy can be recast as
\begin{equation} \label{DWEvarSinvtras}
\delta H_c = - \delta {\bf a} \cdot \mathbf{F} \,,
\end{equation}
where the vector $\textbf{F}$ is defined by the line integral
\begin{equation}
\mathbf{F} = \int_\Gamma ds \, l_z  \mathbf{f}^z + c.c.\,.
\end{equation}
with $ds$ the line element along $\Gamma$, $l_z$ and its c.c. are the components of the covector associated with the vector of the Darboux frame adapted to $\Gamma$ and which is normal to it but tangent to the surface.
 Eq.(\ref{DWEvarSinvtras}) identifies the vector $\textbf{F}$ as the  force acting on the boundary curve $\Gamma$ \cite{interactions}. Furthermore, the quantity $l_z  \textbf{f}^z$ represents the local force acting on the line element $ds$, so that the conserved current associated with translational invariance $\mathbf{f}^z$ is correctly identified as the surface stress tensor.

\vskip1pc\noindent Rotational invariance is a little more involved due to the non-trivial transformation properties of the spinor field. Using the identity (\ref{DWEnormalvect}) for the normal vector $\textbf{n}$ in terms of the spinor field, it is easy to see that the variation in the spinor field induces a variation in $\textbf{n}$ given by
\begin{equation} \label{DWEdeltavecn}
\delta \mathbf{n} = \frac{2}{\lvert \psi \rvert^4} \left( (\bar{\psi}_1 \delta \psi_2 - \psi_2
\delta \bar{\psi}_1)\mathbf{e}_z +(\psi_1 \delta \bar{\psi}_2 - \bar{\psi}_2 \delta
\psi_1)\mathbf{e}_{\bar{z}}\right).
\end{equation}
Making use of expressions (\ref{DWElambdas}) for $\lambda$ and (\ref{DWEdeltavecn}) for $\delta
\textbf{n}$ in the expressions for the currents given above, it is easily seen that $Q$ is given by
\begin{equation}
Q = - i \int dz \wedge d\bar{z} \partial_z \left(\mathbf{f}^z \cdot \delta
\mathbf{X}+\mathbf{c}^z \cdot \delta \mathbf{n}\right) +c.c.\,,
\end{equation}
where $\textbf{c}^{z} = \frac{1}{2 \, \lvert \psi \rvert^2} \frac{\delta
L}{ \delta {\cal V}} \textbf{e}_{\bar{z}}$.

\vskip1pc\noindent Considering now a infinitesimal rotation by a constant angle $\delta
\bm{\omega}$. The embedding functions transform by $\delta \textbf{X} = \delta \bm{\omega} \times
\textbf{X}$; the normal vector also rotates, $\delta \textbf{n} = \delta \bm{\omega} \times
\textbf{n}$. Thus the energy changes by
\begin{equation}
\delta H_c = -i \delta \bm{\omega} \cdot \int dz \wedge d\bar{z}\, \partial_z \mathbf{m}^z+c.c.\,,
\end{equation}
where $\textbf{m}^z= \textbf{X} \times \textbf{f}^z+1/2 \lvert \psi \rvert^{-2} \delta L / \delta {\cal V} \textbf{n} \times \textbf{e}_{\bar{z}}.$
Using an identical argument to the one used to identify the force on the boundary curve $\Gamma$,
we identify the change in the energy associated with a rotation of this curve
\begin{equation}
\delta H_c = - \delta \bm{\omega} \cdot \mathbf{M}
\,,
\end{equation}
where
\begin{equation}
\mathbf{M}= \int_\Gamma ds l_z \mathbf{m}^z + c.c.\,.
\end{equation}
We thus identify $\textbf{M}$ as the total torque acting on this boundary.
$\textbf{m}^z$ and its c.c. are identified as the components of the surface torque tensor
\cite{baltorques}. Using the identity $\mathbf{e}_z = i \mathbf{e}_z \times \mathbf{n}$, 
the component of the torque tensor can
be rewritten in the form
\begin{equation}
\mathbf{m}^z = \mathbf{X} \times \mathbf{f}^z-\frac{i}{2 \, \lvert \psi \rvert^2} \frac{\delta L }
{\delta {\cal V}} \mathbf{e}_{\bar{z}}\,.
\end{equation}

\setcounter{equation}{0}
\renewcommand{\thesection}{Appendix \Alph{section}}
\renewcommand{\thesubsection}{E. \arabic{subsection}}
\renewcommand{\theequation}{E.\arabic{equation}}

\section{Consistency of Euler-Lagrange equations} \label{Compare}

At the end of section \ref{sect2} it was noted that the Canham-Helfrich bending energy differs from the conformally invariant Willmore energy $H_3$, defined by Eq.(\ref{eq:Willmore}) by a topological energy proportional to the Gauss-Bonnet invariant. The corresponding Euler-Lagrange equations therefore coincide; indeed, the corresponding local stresses also coincide. It is instructive to demonstrate this explicitly using the functional form of $H_3$ in terms of $|{\cal A}|$.

\vskip1pc\noindent  In this case derivatives $\partial_z \psi_1$ and $\partial_z \bar{\psi_2}$ appear explicit 
through their Wronskian $\mathcal{A}$ and its c.c..
The component $f^{z}_{\phantom{z}\bar{z}}$ given by 
Eq.(\ref{DWEFZzb})  is replaced by 
\begin{equation} \label{DWEFZzb1}
f^{z}_{\phantom{z}\bar{z}} \to f^{z}_{\phantom{z}\bar{z}}
+ {\cal V} \left(\psi_2 \frac{\partial L}{\partial (\partial_z \psi_1)} - \bar{\psi}_1 \frac{\partial L}{\partial (\partial_z \bar{\psi}_2)}\right) \,.\end{equation}

\vskip1pc \noindent  For a functional which depends on derivatives of $\psi$ no higher than first, the relevant components of the stress tensor are
\begin{subequations}\begin{eqnarray}
f^z_{\phantom{z}z}  &=& -\frac{1}{2} \left(\psi_1 \left(\frac{\partial L}{\partial
\psi_1}-\partial_z \left(\frac{\partial L}{\partial (\partial_z \psi_1)}\right)\right)+\bar{\psi}_2 \left( \frac{\partial L}{\partial \bar{\psi}_2}-\partial_z\left(\frac{\partial L}{\partial(\partial_z \bar{\psi}_2)}\right)\right)\right),\\
f^{z}_{\phantom{z}\bar{z}} &=& \frac{\bar{{\cal A}}}{2 \, \lvert \psi \rvert^2} \frac{\partial L}
{\partial {\cal V}}+{\cal V} \left(\psi_2 \frac{\partial L}{\partial (\partial_z \psi_1)} - \bar{\psi}_1 \frac{\partial L}{\partial (\partial_z \bar{\psi}_2)}\right),\\
f^z & = & -\partial_{\bar{z}} \left(\frac{1}{2 \lvert \psi \rvert^2}
\frac{\partial L}{\partial {\cal V}} \right) + \frac{1}{\lvert \psi \rvert^2 } \left(\psi_2 \left(\frac{\partial L}{\partial \psi_1}-\partial_z\left(\frac{\partial L}{\partial (\partial_z {\psi}_1)}\right)\right) -\bar{\psi}_1 \left( \frac{\partial L}{\partial \bar{\psi}_2} -\partial_z\left(\frac{\partial L}{\partial (\partial_z \bar{\psi}_2)}\right)\right) \right)\,.
\end{eqnarray}
\end{subequations}
Note that for functionals involving higher order derivatives, the corresponding functional derivatives will contain
extra terms arising from integration by parts. 

\vskip1pc \noindent 
The required partial derivatives are
\begin{subequations}
 \begin{align}
\frac{\partial L}{\partial \psi_1} & = -\frac{4}{|\psi|^4}
\left(\bar{\mathcal{A}}\partial_z \bar{\psi}_2 + 2
\frac{|\mathcal{A}|^2}{|\psi|^2}\bar{\psi}_1\right), &\quad \frac{\partial L}{\partial(\partial_z
\psi_1)} & = \frac{4}{|\psi|^4}\bar{\mathcal{A}} \bar{\psi}_2,\\
\frac{\partial L}{\partial \bar{\psi}_2} & = \frac{4}{|\psi|^4}
\left(\bar{\mathcal{A}}\partial_z \psi_1 -2
\frac{|\mathcal{A}|^2}{|\psi|^2}\psi_2\right), &\quad \frac{\partial L}{\partial(\partial_z
\bar{\psi}_2)} & = -\frac{4}{|\psi|^4}\bar{\mathcal{A}} \psi_1\,.
\end{align}
\end{subequations}
Making use of the Codazzi-Mainardi Eq.(\ref{DWECM}) we have that the corresponding functional  derivatives are
\begin{equation}
\frac{\delta \tilde{L}}{\delta \psi_1} = - \bar{\psi}_2 \partial_{\bar{z}} K, \qquad \frac{\delta
\tilde{L}}{\delta \bar{\psi}_2} = \psi_1 \partial_{\bar{z}} K, \qquad \frac{\delta \tilde{L}}{\delta
{\cal V}} = 0\,.
\end{equation}
Substituting these identities into expressions (\ref{DWEFZz}) and (\ref{DWEFZ}) for the components
of the stress tensor reproduces a stress identical to that for the Canham-Helfrich energy written in
Eq.(\ref{DWEfcmpWlmr}). It is worth pointing out that the two stress tensors did not need to
coincide: they could have differed by a null stress.

\end{appendix}

\end{document}